%Paper: hep-th/9306088
%From: udah130@elm.cc.kcl.ac.uk
%Date: Fri, 18 Jun 1993 14:06:20 EDT

\input phyzzx

\Pubnum{ \vbox{ \hbox{UM-P-93/38} \hbox{KCL-93-7} \hbox{9306088}} }
\pubtype{}
\date{May 1993}

\titlepage

\title{The Space of Solutions of Toda Field Theory}

\author{G. Papadopoulos}
\address{Department of Mathematics \break King's College London\break
         London WC2R 2LS}
\andauthor{B. Spence}
\address{School of Physics\break University of Melbourne\break Parkville
          3052 Australia}

\abstract{A new parameterisation of the solutions of Toda field theory is
introduced. In this parameterisation, the solutions of the field equations are
real, well-defined functions on space-time, which is taken to be
two-dimensional Minkowski space or a cylinder. The global structure of the
covariant phase space of Toda theory is examined and it is shown that it is
isomorphic to the Hamiltonian phase space.  The Poisson brackets of Toda theory
are then calculated.  Finally, using the methods developed to study the Toda
theory, we extend these results to the non-Abelian Toda field theories. }
\endpage

%%%%%%%%%%%%%%%%%%%%%%%%%%%%%%%%%%%%%%%%%%%%%%%%%%%%%%%%%%%%%%
\def\half{{1\over2}}

\def\pbr#1#2{ \{#1,#2\} }
\def\pl{Phys. Lett.\ }

\def\np {Nucl. Phys.\ }

\def\exp{{\rm exp}}

\def\dpl{\partial_+}
\def\dmi{\partial_-}

\def\intx{\int_0^1\!dx\,}
\def\l {\lambda}
\def\G{{\cal L}(G)}

\def\P{{\cal L}(P)}
\def\Gp{{\cal L}^+(G)}
\def\Gm{{\cal L}^-(G)}
\def\Gep{{\cal L}^+_{\epsilon}(G)}
\def\Gem{{\cal L}^-_{\epsilon}(G)}
\def\Geo{{\cal L}^{{}^0}_{\epsilon}(G)}
\def\a{\alpha}

\def\H{{\cal H}}

\def\Cab{C_{ij}}
\def\Kab{K_{ij}}
\def\nua{\nu^i}
\def\mua{\mu^i}
\def\pa{\phi^i}
\def\pb{\phi^j}
\def\Ea{E_\alpha}
\def\Ema{E_{-\alpha}}
\def\Sumap{\sum_{\alpha\in\Phi^+}}

\def\Sumd{\sum_{i}}
\def\lra{\big\vert\lambda_i\big>}
\def\lla{\big<\lambda_i\big\vert}

\font\mybb=msbm10 at 12pt
\def\bb#1{\hbox{\mybb#1}}
\def\RN {\bb{R}}

\REF\toda {M. Toda, Phys. Rep. {\bf 18C} (1975) 1.}
\REF\ls {A.N. Leznov and M.V. Saveliev, Lett. Math. Phys. {\bf 3} (1979)
      489; Commun. Math. Phys. {\bf 74} (1980) 111.}
\REF\pma {P. Mansfield, \np {\bf B208} (1982) 277; {\bf B222} (1983) 419.}
\REF\ot {  D. Olive and N. Turok, \np {\bf B220} (1983) 491.}
\REF\gn   { J.-L. Gervais and A. Neveu, \np {\bf B224} (1983) 329.}
\REF\thorn {E. Braaten, T. Curtright, G. Ghandour and C. Thorn, Phys. Lett.
{\bf 125B} (1983) 301.}
\REF\bab {O. Babelon, Phys. Lett. {\bf 215B} (1988) 523}
\REF\babtwo {O. Babelon, F. Toppan, and L. Bonora, Commun. Math. Phys. {\bf
140} (1991) 93.}
\REF\gervais {J-L Gervais, \lq\lq Recent Progress of The Liouville Approach to
2D Gravity and its Toda (W) Generalisations", LPTENS-92/36.}
\REF\bg {A. Bilal and J-L Gervais, Phys. Lett. {\bf 206B} (1988) 412; Nucl.
Phys. {\bf B314} (1989) 646; {\bf B318} (1989) 579.}
\REF\ms{P. Mansfield and B. Spence, \np {\bf B362} (1991) 294.}
\REF\nonsttt{L. Feh\'er, L. O'Raifeartaigh, P. Ruelle, I. Tsutsui and A. Wipf,
         Phys. Rep. {\bf 222}, No 1 (1992) 1.}
\REF\sato {M. Sato, RIMS Kokyuroku {\bf 439} (1981) 30; E. Date, M. Jimbo, M.
Kashiwara and T. Miwa, \lq\lq Transformation Groups for Soliton Equations" in
Proc. of RIMS Symposium on Non-linear Integrable Systems-Classical Theory and
Quantum Theory, World Scientific Publ. Co. Singapore (1983).}
\REF\irish{P. Forg\'acs, A. Wipf, J. Balog, L. Feh\'er and
           L. O'Raifeartaigh, \pl {\bf 227B} (1989) 214.}
\REF\usthree{G. Papadopoulos and B. Spence, {\it A Covariant Canonical
     Formulation of Liouville Field Theory}, hepth/9303076, \pl B, in press.}
\REF\us{G. Papadopoulos and B. Spence, \pl {\bf 295B} (1992) 44.}
\REF\ustwo{G. Papadopoulos and B. Spence, {\it The Canonical Structure of
            the Wess-Zumino-Witten Model}, to appear in the Proceedings of
             the NATO ASI Conference ``Low Dimensional Topology and
    Quantum Field Theory'', Newton Institute for Mathematical Sciences,
    Cambridge, September 1992.}
\REF\lstwo {A.N. Leznov and M.V. Saveliev, Lett. Math. Phys. {\bf 6}(1982) 505;
Commun. Math. Phys. {\bf 83} (1983) 59; J. Sov. Math. {\bf 36} (1987) 699; Acta
Appl. Math. (1989) 1.}
\REF\nonst{L. O'Raifeartaigh and A. Wipf, \pl {\bf 251B} (1990) 361.}
\REF\nonstt{F.A. Bais, T. Tjin and P. van Driel, \np {\bf B357} (1991) 632.}
\REF\witten {E. Witten, Commun. Maths. Phys. {\bf 92} (1982) 455.}
\REF\ellis {S.W. Hawking and G. F. R. Ellis, {\it The large scale structure of
   space-time}, CUP (1973).}
\REF\bersh{M. Bershadsky, Commun. Math. Phys. {\bf 139} (1991) 71.}

\sequentialequations

%%%%%%%%%%%%%%%%%%%%%%%%%%%%%%%%%%%%%%%%%%%%%%%%%%%%%%%%%%%%%%%

\chapter{Introduction}
Toda theory [\toda -\thorn] is a conformal field theory and provides a popular
arena
for the investigation of general features of two-dimensional integrable systems
such as quantum groups, W-algebras and integrable hierarchies (see for example
refs. [\bab -\sato]  and
references therein). The study of some of these properties utilises an explicit
parameterisation of the solutions of the classical field equations of Toda
theory and the associated Poisson bracket algebra of these parameters.
Parameterisations of the solutions of the field equations of Toda theory were
given in refs. [\ls, \pma, \irish].  These parameterisations describe the
solutions of Toda theory on a two-dimensional {\sl Minkowski} space-time.  The
difficulty with them is that the ranges of the associated parameters are not
specified.  Because of this, the solutions  are not well-defined functions over
all space-time (they are infinite at certain regions of space-time) and/or the
solutions become complex for some choices of these parameters.  For a detailed
discussion of these problems in the case of Liouville theory see ref.
[\usthree].  A related problem is to find  the relationship between the
parameters of the solutions  and the initial data of Toda theory, and define
the covariant phase space and Poisson brackets of the theory.  To our
knowledge, these problems have not been resolved in the literature.

In a previous paper [\usthree], we presented a new parameterisation of the
solutions of the field equations of Liouville theory.  In this
parameterisation, the solutions of the field equations are real, well-defined
functions over all of  space-time.  Furthermore, we showed that there is a
diffeomorphism that relates the independent parameters of the solutions to the
initial data of the theory, and we used this diffeomorphism to prove that the
covariant and Hamiltonian phase spaces of this theory are diffeomorphic as
symplectic manifolds.  Finally, we proved that our parameterisation of the
solutions of Liouville theory can be obtained, by using the Hamiltonian
reduction methods of ref. [\irish],  from the parameterisation of the solutions
of Wess-Zumino-Witten (WZW) theory of refs. [\us,\ustwo].

In this paper, we will extend these results on Liouville theory to Toda theory.
 In particular, we will show that there is a parameterisation of the solutions
of the field equations of Toda theory such that these solutions are real
functions over all of the two-dimensional space-time, which is taken to be
either a {\sl cylinder} or {\sl Minkowski} space.  We will also construct an
explicit diffeomorphism that relates the independent parameters of the
solutions to the initial data of the theory and we will use this diffeomorphism
to construct a symplectic diffeomorphism between the Hamiltonian and covariant
phase spaces of Toda theory. To achieve these results, we will employ the
Hamiltonian reduction method of ref. [\irish] and the parameterisation of refs.
[\us, \ustwo] of the solutions of the WZW model.  The Poisson brackets of Toda
theory are induced from those of the WZW models, and we show that these
brackets are associated to the co-tangent bundle of a loop space which arises
naturally in this reduction.  Finally, we will describe how these results can
be extended to treat the non-Abelian Toda theories [\lstwo] that can be
obtained using the non-standard Hamiltonian reduction methods of refs. [\nonst,
\nonstt].

%%%%%%%%%%%%%%%%%%%%%%%%%%%%%%%%%%%%%%%%%%%%%%%%%%%%%%%%%%%%%%%%%%%%%%%%%%%

\chapter{Toda Field Theory and Hamiltonian Reduction}

Let $\G$ be a Lie algebra of a (semi)simple group $G$ and $\H$ be a Cartan
subalgebra of $\G$.  We introduce a Chevalley basis $(H_i, E_{\a^i},
E_{-\a^i})$ in  $\G$, where $\Delta\equiv \{\a^i, i=1, \dots,l={\rm rank}  \G
\}$ is the space of simple roots, $H_i\equiv {2\a^i\cdot H \over
\vert\a^i\vert^2}$, $H\in \H$, $E_{ \pm\a^i}$ are the step vectors of the
simple roots and $[H_i, H_j]=0$, $[E_{\a^i}, E_{-\a^i}]= H_i$ and $[H_i, E_{\pm
\a^j}]=\pm K_{ji}  E_{\pm \a^j}$ (with no summation over $j$).  The matrix
$K\equiv \{K_{ij}\}$ is the Cartan matrix of $\G$, i.e. $K_{ij}= {2
\a^i\cdot\a^j \over \vert\a^j\vert^2}$.  The symbols $\Phi^+$ ($\Phi^-$) will
denote the sets of positive (negative) roots, respectively, and $\Phi\equiv
\Phi^+\cup \Phi^-$ is the space of all roots of $\G$. We will also use the
symbols $\Gp$ and $\Gm$ to denote the sets of step vectors along the positive
and negative roots respectively.  Finally, we normalise the Killing form as
follows: $\Tr(H_i\cdot H_j)
=C_{ij}$, $\Tr(E_{\a^i}\cdot E_{-\a^j}) = {2\over\vert\a^i\vert^2}\delta_{ij}$
and $\Tr(E_{\a^i}\cdot H_j) = 0$, where $C_{ij}\equiv {2\over\vert\a^i\vert^2}
K_{ij}$.

The Lagrangian of Toda field theory is
$$
L=-{k\over 8\pi} \Big(C_{ij}\partial_+\phi^i\dmi\phi^j - \Sumd M^i
        \exp\Big(\Kab\pb\Big)\Big),
\eqn\one
$$
where $\phi$ is a map from a
cylinder $S^1\times \RN$ to $\RN^l$ and $k$, $M^i$ are real, {\sl non-zero}
coupling constants. The pairs
$(x,t): 0\leq x<1, -\infty<t<\infty$ are the
co-ordinates of $S^1\times \RN$ and
$x^\pm = x \pm t, \partial_\pm = \half(\partial_x \pm\partial_t)$.
The equations
of motion following from the Lagrangian \one\ are
$$
     \dpl\dmi\pa + {\vert\a^i\vert^2\over 4} M^i \exp\Big(\Kab\pb\Big)=0.
       \eqn\two
$$

In ref. [\irish] it was shown that Toda theory can be derived, using
Hamiltonian reduction, from the WZW model with target space the group $G$,
which is taken to be the maximal non-compact real Lie group associated to the
Lie algebra $\G$.  An important property of this group is that it admits {\sl
locally} a Gauss decomposition. The field equations of the WZW model are
$$
   \partial_-(\partial_+  g\ g^{-1})=0, \eqn\four
$$
where $g$ is a map from a two-dimensional space-time, a cylinder $S^1\times
\RN$,  to  the group $G$ and the conserved currents (subject to field
equations) are
$$
  \eqalign{
J_+(\l) &=-\kappa\Tr(\l\cdot\partial_+  g\ g^{-1}),\cr
         J_-(\l) &= \kappa\Tr(\l\cdot g^{-1}\partial_- g), \cr}  \eqn\six
$$
where $\kappa ={k\over 4\pi}$, $k$ is the coupling constant, and
$\l$ is an arbitrary element of the Lie algebra $\G$.

The WZW model reduces to Toda theory upon imposing the first class constraints
$$
 \eqalign{
J_+(E_{\a^i})&= \kappa \mua, \quad \a^i\in \Delta, \qquad J_+(E_{\a})=0, \quad
\a\in \Phi ^+-\Delta, \cr
  J_-(E_{-\a^i})&= -\kappa \nua, \quad \a^i\in \Delta, \qquad  J_-(E_{-\a})=0,
\quad
\a\in \Phi ^+-\Delta,}
 \eqn\seven
$$
where the components of $\mu$ and $\nu$ are real, {\it non-zero} constants. The
authors of ref. [\irish] solved the above constraints using the Gauss
decomposition properties of the group $G$ and parameterised the solutions of
Toda theory using the parameterisation of solutions of the WZW model of ref.
[\witten].  This parameterisation of the solutions of Toda theory was then
related to that given in ref. [\ls].

%%%%%%%%%%%%%%%%%%%%%%%%%%%%%%%%%%%%%%%%%%%%%%%%%%%%%%%%%%%%%%%%%%%%%%%%

\chapter { A Parameterisation of the Toda Solutions}
To give a new parameterisation of the space of solutions of the field equations
of Toda theory,
eqn. \two, we will use the Hamiltonian reduction method
\foot {Here we will not treat the Hamiltonian reduction method as fundamental
but rather as a device to construct a parameterisation of the solutions of Toda
theory. This is because of difficulties in dealing with global issues in
Hamiltonian reduction. This does not affect the results of this paper.}
as described above and the parameterisation
$$
\eqalign{   g(x,t) &= U(x^+) {\cal W}(A;x^+,x^-) V(x^-), \crr
                 {\cal W}(A;x^+,x^-) &=
                     P\,\exp\!\int_{x^-}^{x^+}\!A(s)ds,\cr } \eqn\three
$$
of the solutions of the WZW model (on a cylinder) given in ref. [\us]. The maps
$U$ and $V$ are periodic maps from the real line ${\RN}$ into the
group $G$, and the field $A$ in the path-ordered exponential is a
 $\G$-valued periodic one-form on
the real line. The expression for $g(x,t)$ in eqn. \three\ is then
periodic in $x$ and solves the field equations \four.
The parameterisation \three\ has the symmetry
$$
     \eqalign { & U(x)\rightarrow U(x)h(x),
          \quad V(x)\rightarrow h^{-1}(x)V(x),
         \cr &A(x)\rightarrow -h^{-1}(x)\partial_xh(x) +
           h^{-1}(x)A(x)h(x),     \cr}       \eqn\five
$$
where $h$ is an element of the loop group of $G$.

 Rewriting the constraints \seven\ in terms of the variables $U, V, A$ of the
parameterisation \four, they become
$$
   \eqalign{(\nabla_+U U^{-1})(E_{\a^i}) \equiv
  \big(\partial_+U U^{-1}+ UA(x^+)U^{-1}\big)(E_{\a^i}) &= \kappa\mua,\cr
   (V^{-1}\nabla_-V)(E_{-\a^i})\equiv
    \big(V^{-1}\partial_-V -V^{-1}A(x^-)V\big)(E_{-\a^i}) &= -\kappa\nua,\cr}
    \eqn\eight
$$
and {\sl zero} along the directions of the remaining step vectors of the
negative and positive roots respectively.
To solve these constraints, we recall that the group $G$ admits locally a Gauss
decomposition, $G=DBC$, where the Lie algebras of the subgroups $D$ and $C$ of
$G$ are $\Gp$ and $\Gm$ correspondingly. The subgroup $B$ of $G$ is a maximal
torus of $G$ with Lie algebra $\H$.  The parameters $U,V$ can then be
decomposed as
$$
  U = d_{{}_L}b_{{}_L} c_{{}_L}, \quad V = d_{{}_R} b_{{}_R}c_{{}_R},
\eqn\twelve
$$
where
$$
   \eqalign{  d_{{}_L} &= \exp\Big(\Sumap x_L^\a\Ea\Big),
     \quad c_{{}_L} = \exp\Big(\Sumap y_L^\a\Ema\Big),
    \crr   b_{{}_L} &= \exp\Big(\Sumd \pa_LH_i\Big).    \cr}   \eqn\thirteen
$$
and similarly for $V$.
The constraints \eight\ can be rewritten in terms of  $d_{{}_R}$, $\phi_L$,
$\phi_R$ and $c_{{}_L}$  as follows:
$$
     \eqalign{ \nabla_+ c_{{}_L}\; c_{{}_L}^{-1}\vert_{\Gm} &= \Sumd
{\vert\a^i\vert^2\over2}\mua\  E_{-\a^i}
                \exp\big(\Kab\pb_L\big), \crr
   d_{{}_R}^{-1}\nabla_- d_{{}_R}\vert_{\Gp} &=
\Sumd {\vert\a^i\vert^2\over2}\nua\  E_{\a^i}
               \exp\big(\Kab\pb_R\big). \cr}
\eqn\fourteen
$$
where $\nabla$ is the covariant derivative with respect to the connection $A$
as defined in eqn. \eight.

The solutions of the Toda equations of motion can be parameterised in terms of
$\phi_L$, $\phi_R$, $c_{{}_L}$, $d_{{}_R}$ and $A$.  To do this, we perform a
Gauss decomposition for the WZW field $g$;
$$
     g = dbc,          \eqn\nine
$$
where
$$
    \eqalign{ d &= \exp\Big(\Sumap x^\a\Ea\Big),
     \quad c = \exp\Big(\Sumap y^\a\Ema\Big),
    \crr   b &= \exp\Big(\Sumd \pa H_i\Big).    \cr}   \eqn\ten
$$
The Toda field is identified as the map $\phi$ that appears in the definition
of $b$ in the above equation.  To project the Toda field from the expression
\nine\ for $g$, we use the standard method of the $l$ normalised lowest weight
states $\lra$ of finite
dimensional representations of $\G$ (with $H_j\lra = -\delta_{ij}\lra)$.
{}From eqns. \nine, \three\ and $<\l_i|g|\l_i>={\rm exp}(- \phi^i)$,
we get
$$
    \exp\big(-\pa(x,t)\big) = \exp\big(-\pa_L(x^-) -
                 \pa_R(x^+)\big)\lla
                          c_{{}_L}(x^+) \,W(A;x^+,x^-) d_{{}_R}(x^-) \lra.
     \eqn\fifteen
$$
It is straightforward to prove that $\phi$ in eqn. \fifteen\ satisfies the Toda
equation provided that the parameters $c_{{}_L}$ and $d_{{}_R}$ satisfy eqns.
\fourteen\ and $M^i=\vert \a^i\vert^2 \mu^i \nu^i$.

To find the independent parameters of the solutions of the Toda equations, we
have still to gauge-fix the symmetry \five\ and solve the constraints
\fourteen.  The transformation \five\ can be gauge-fixed by setting $U=e$,
where $e$ is the identity element of the group $G$.  This implies that
$\phi_L=0$ and $c_{{}_L}=e$.  Then the first constraint in eqn. \fourteen\ can
be easily solved by taking
$$
       A^{-\a^i} = \mua, \quad \alpha^i \in \Delta; \qquad A^{-\a}=0, \quad
\a\in \Phi^+-\Delta          \eqn\seventeen
$$
where $A^{-\a}=\Tr(\Ea A)$, $\a\in \Phi^+$.  The transformations
$$
     \eqalign {d_{{}_R}(x)&\rightarrow u^{-1}(x)d_{{}_R}(x),\cr
       A(x)&\rightarrow -u^{-1}(x)\partial_x u(x) + u^{-1}(x)A(x)u(x), \cr}
    \eqn\eighteen
$$
where $u$ is an element of the loop group of the group $D$, preserve the
solution \seventeen\ and the second constraint of eqn. \fourteen.  We can use
this residual symmetry to set $d_{{}_R}=e$.  In this gauge, the second
constraint of eqn.\fourteen\ can be solved as
$$
     A^{\a^i} = -\nua\exp(\Kab\pb_R)\quad \alpha^i \in \Delta; \qquad A^{\a}=0,
\quad \a\in \Phi^+-\Delta  ,     \eqn\twenty
$$
where $A^\a = \Tr(\Ema A)$, $\a \in \Phi^+$.  To complete the discussion of the
independent parameters of the solutions of the Toda equation, we observe that
the constraints are independent from the parameter $c_{{}_R}$ and so without
loss of generality we can set $c_{{}_R}=e$.  This can be understood as fixing
an associated symmetry in the space of parameters.

Thus the {\sl independent} parameters of the solutions of Toda theory are the
variables $\phi_R$ and the components $a^i$ of the connection $A$ along the
directions of the Cartan subalgebra $\H$ of $\G$.
The solutions of the Toda theory parameterised in terms of these independent
variables are
$$
     \exp\bigg(-\pa(x,t)\bigg) =
       \exp\bigg(-\pa_R(x^-)\bigg)\lla W(A;x^+,x^-) \lra,    \eqn\tone
$$
where the components of the connection $A$ corresponding to the subspaces
$\Gem$ and $\Gep$  are given in eqns. \seventeen,  \twenty\ and the components
$a^i$ along the directions of the Cartan subalgebra $\H$ of $\G$ are {\sl
unrestricted}.  The above  solutions $\phi$ are periodic in $x$ whenever
$\phi_R$ and $a$ are periodic functions on the real line.  The proof that the
expression for $\phi$ in eqn. \tone\ solves the Toda field equations follows
from its method of construction.  Alternatively this can be shown directly.

%%%%%%%%%%%%%%%%%%%%%%%%%%%%%%%%%%%%%%%%%%%%%%%%%%%%%%%%%%%%%%%%%%%%%%%%%%%%

\chapter{The  Phase Space Structure of Toda Theory}

The Lagrangian symplectic form $\Omega$ of a field theory is
defined as the integral over a Cauchy surface of the time component of the
symplectic current $S^{\mu}=\delta \phi^I \delta \big(\partial
L/\partial(\partial_{\mu}\phi^I)\big)$, where $\phi$ is the field
and $L$ is the Lagrangian of a theory.  This two-form is closed and independent
of the choice of the Cauchy surface that we have used to define it. The
covariant phase space $P_C$ of a
theory is then defined as the space of parameters of the solutions of the
Lagrangian equations of motion of the system, equipped with the Lagrangian
symplectic form given in terms of these parameters.

The Lagrangian symplectic form of the Toda theory is
$$
    \Omega = {k\over 16\pi} \intx\Cab\,\delta\pa\,\partial_t\delta\pb,
\eqn\sixteen
$$
evaluated at $t=0$, where the map $\phi$ in eqn. \sixteen\ satisfies
the Toda field equations of motion \two. We wish to insert into the form
$\Omega$
the Toda solution \tone, to express $\Omega$ in terms of the independent
parameters of the solutions. Indeed, substituting the solution \tone\ into the
symplectic form \sixteen\ we get
$$
   \Omega = -{k\over 16\pi}\intx \Cab\big(\delta\pa_R\,\partial_x\delta\phi^j_R
-2\delta\pa_R\,
        \delta a^j \big).      \eqn\tone
$$
where $a^i=-<\l_i|a|\l_i>$.  The Poisson brackets are easily obtained from
this, and are
$$
  \eqalign{ \pbr{\pa_R(x)}{\pb_R(y)} & = 0, \crr
          \pbr{\pa_R(x)}{a^j(y)} & ={8 \pi\over k}
(C^{-1})^{ij}\delta(x,y),\crr
          \pbr{a^i(x)}{a^j(y)} & = {8 \pi\over k}(C^{-1})^{ij}
                                     {\partial_x}\delta(x,y).\crr}
     \eqn\ttwo
$$
Note that these brackets are the Poisson brackets
on the co-tangent bundle of the loop space of   $\RN^l$.  Finally, it can be
easily seen that the Hamiltonian reduction described in the previous section
relates the Poisson brackets of eqn. \ttwo\ to the Poisson brackets of the WZW
model given in refs. [\us, \ustwo].

There are symplectic diffeomorphisms amongst the covariant phase space, the
Hamiltonian phase space and the space of initial data of Toda theory.  It is
straightforward to prove that the latter two spaces are isomorphic.  In the
following we will use the parameterisation \tone\ of the covariant phase space
of Toda theory to construct a symplectic diffeomorphism between the covariant
phase space and the space of initial data of Toda theory.  The space of initial
data is the space of Toda fields $\phi$ and their time derivatives
${\partial\pa\over\partial t}$ at $t=0$. The space of initial data can be
equipped with the symplectic form induced from eqn. \sixteen.
{}From the Toda solution \tone, we find immediately
that the map between the initial data $f^i(x) \equiv \pa(x,0)$, $w^i(x)
\equiv {\partial\pa\over\partial t}(x,0)$ and our covariant phase space
parameters $\phi_R, a^i$ is
$$
   \eqalign { f^i(x) &= \pa_R(x), \cr
            w^i(x)  &= - \partial_x\pa_R(x)+2 a^i(x). \cr}  \eqn\tthree
$$
This map is clearly a diffeomorphism. Moreover it is a symplectic
diffeomorphism from the space of initial data to the space of parameters of the
solutions of the Toda theory because,   as
it is easy to show, it maps the symplectic form \tone\ into the
form $\Omega = {k\over 16\pi}\intx \Cab\delta f^i(x)\,\delta w^j(x)$.

To show that the solutions of the Toda theory parameterised in terms of $a$ and
$\pa_R$ are {\it real}, we can repeat the same arguments as those used in ref.
[\usthree] for the solutions of the Liouville theory.  Indeed,
for solutions analytic in the time co-ordinate $t$, this
can be shown explicitly by expressing these solutions as power series in $t$
and observing that the solution $\phi$ is real at every order in the expansion.
For solutions $\phi$ which are not analytic in $t$, we note that the Toda
theory has a
well-posed initial value problem [\ellis], which implies
that there is a unique solution of the Toda equation for each pair $(f,w)$ of
initial data.  Now we observe that if $\phi$ is a solution of the
Toda equation and $\phi$ is complex, then its complex conjugate ${\phi}^*$
is a solution as well.  However, if the initial data of $\phi$ are
real, it is easy to prove that both $\phi$ and ${\phi}^*$ have the same initial
data.  Hence they must be equal ($\phi={\phi}^*$), and $\phi$ is real.

%%%%%%%%%%%%%%%%%%%%%%%%%%%%%%%%%%%%%%%%%%%%%%%%%%%%%%%%%%%%%%%%%%%%%

\chapter {Non-Abelian Toda Theories}

The methods developed in the previous sections can also be used to study the
space of solutions of the models constructed from the WZW model  with
non-standard Hamiltonian reductions. These models are the non-abelian Toda
theories of refs. [\nonsttt, \nonst, \nonstt].
As this is a relatively straightforward
generalisation of the above discussion of ordinary Toda theory, we will only
sketch
the main steps here. In the following, we will only consider the cases with
first class constraints
\foot {Toda systems with second class constraints involve the inclusion of free
fermions [\bersh]  and can be incorporated in our construction in a
straightforward way.}.

As in the case of Toda theory, we begin with a WZW model with target space a
semi-simple group $G$ and introduce a vector $\epsilon$ on the space of
fundamental co-weights,  i.e. the space that is dual to the space of simple
roots of the Lie algebra $\G$ of the group $G$.  The vector $\epsilon$ has
components that are either one or  zero in the standard basis $\{\Lambda_i\}$
($\a^j \cdot \Lambda_i= \delta^j_i$) of this space.  Then a gradation can be
introduced on $\G$ with respect to $\epsilon$ as follows: A generator $T$ of
$\G$ has grading $n(T)$ provided that
$$
[H_\epsilon, T]= n(T) T
\eqn\done
$$
where $H_\epsilon\equiv\epsilon\cdot H$, $H\in \H$, and $n(T)$ is an integer.
In the following, we denote the grading of a generator with a superscript, i.e.
$T^n$ means that $T$ has grading $n(T)$ in the above gradation.   We can use
the above gradation to decompose the Lie algebra $\G$ into the subalgebras
$\Gep\equiv \{T\in \G: n(T)>0\}$, $\Gem\equiv \{T\in \G: n(T)<0\}$ and
$\Geo\equiv \{T\in \G: n(T)=0\}$.   Notice that for the generators of the
Cartan subalgebra $\H$ of $\G$, the grading is zero so that $\H\subset \Geo$;
for the step vectors associated to the positive roots the grading is a positive
integer $n(E_\a)\geq 0$  and  for the step vectors associated to the negative
roots the grading is a negative integer $n(E_{-\a)}\leq 0$.  Next we define $P$
to be the subgroup of $G$  generated by the those generators of $\G$ that have
grading {\sl zero}.  It is clear that $\P\equiv\Geo$ and contains the Cartan
subalgebra $\H$ of $\G$.  Following ref. [\nonst], we introduce the constraints
$$
J_+\vert_{\Gem}=M,   \qquad J_-\vert_{\Gep}=N,
\eqn\dtwo
$$
on the currents $J_+, J_-$ of the WZW model with target space the group $G$,
where  $M\in \Gem$ and $N\in \Gep$ are constant vectors. The constant vectors
$M$ and $N$  are  chosen such that their components are {\sl non-zero}  along
the directions of the subspaces of $\Gem$ and $\Gep$  with grading $n$ equal to
$-1$ and $+1$ respectively, and their remaining components {\sl vanish}.

The Lagrangian of these models is given by $L=L_{WZW}+V_{pot.}$ where $L_{WZW}$
is the Lagrangian of a WZW model  with target space the subgroup $P$ of $G$ and
$V_{pot.}$ is a potential term given by
$V_{pot.}={\rm{Tr}} (M p N p^{-1})$ where $p$ is a map from $S^1\times \RN$
into  $P$.  The associated field equations are
$$
\partial_-(\partial_+p p^{-1})+ [p N p^{-1},M]=0.
\eqn\ddtwo
$$

To solve the above constraints we insert our parameterisation of the solutions
of the
WZW model, eqn. \three, into these constraints, and gauge-fix $U = e$ as in the
case of Toda theory in section three.
In this gauge, the first constraint in eqn. \dtwo\  becomes
simply the requirement that the corresponding components of the
connection $A$ of the WZW solution \three\
be fixed to be constants, i.e.
$$
A\vert_{\Gem}=M.
\eqn\dthree
$$
To solve the second constraint in eqn.\dtwo,  we perform a local decomposition
of the group $G$ with respect to its subgroup $P$, $G=SPQ$, where $S$ and $Q$
are the subgroups of $G$ generated by the generators of $\G$ with positive and
negative $n$ grading respectively, i.e. ${\cal L}(S)\equiv\Gep$ and ${\cal
L}(Q)\equiv\Gem$
\foot {In the case where $\epsilon =\sum_{i} \Lambda_i$, $i=1, \dots, {\rm
rank} \G$, we recover the usual Toda case, with  $P=B$, $S=D$ and $Q=C$.}.
Then the parameter $V$ of the solutions of the WZW model can be decomposed as
$V=s_{{}_R}p_{{}_R}q_{{}_R}$ where $s_{{}_R}\in S$, $p_{{}_R}\in P$ and
$q_{{}_R}\in Q$.   Next we observe  that even after imposing the gauge $U=e$,
the constraints \dtwo\ remain invariant under the action of the gauge
transformations of eqn. \five\ provided that the parameter $h$ of these gauge
transformations is restricted to be an element of the loop group of the group
$Q$.  We can use this residual symmetry of the solution \dthree\ to set
$q_{{}_R}$ in the decomposition of the parameter $V$ equal to the identity
element of $G$, and express the solution of the second constraint of eqn.
\dtwo\ in terms of the connection $A$ and $p_{{}_R}$ as follows:
$$
A\vert_{\Gep}=-p_{{}_R}N p_{{}_R}^{-1}.
\eqn\dfour
$$
{}From eqns. \dthree\ and \dfour\ it is clear that the {\sl independent}
parameters that parameterise the solutions of the non-Abelian theories are $a$
and $p_{{}_R}$, where   $a\equiv A|_{\P}$ is the  restriction of the connection
$A$ along the directions of the Lie algebra of the subgroup $P$ of $G$, and the
parameters $p_{{}_R}$ appear in the decomposition $V=s_{{}_R}p_{{}_R}q_{{}_R}$
of $V$ above.

The fields of the non-abelian Toda theory are identified with the $p$ component
in the decomposition $g=spq$ of the field $g$ of the associated WZW theory.
Next, using the independent parameters $a$ and $p_{{}_R}$, we can parameterise
the most general solution of the field equations \ddtwo\ of these models.
Indeed from eqns \three, \dthree\ and \dfour, we get
$$
p(x,t)= W(A;x^+,x^-)\vert_{{}_P}\, p_{{}_R}(x^-),
\eqn\dfive
$$
where $W(A;x^+,x^-)\vert_{{}_P}$ is the $P$-component of $W(A;x^+,x^-)$ in the
decomposition $G=SPQ$ of the group $G$, and $A$ is the connection with
components along the directions of ${\cal L} (S)$ and ${\cal L} (Q)$ given by
the eqns. \dthree\ and \dfour\ respectively.  The components of $a$, $a\equiv
A|_{\P}$, are unrestricted and together with $p_{{}_R}$ are the independent
parameters of the solutions of the theory.  Finally, the periodicity of $p$ in
the co-ordinate $x$ follows from the periodicity of the parameters $a$ and
$p_{{}_R}$.

The Poisson brackets of the parameters $ p_{{}_R}$ and $a$  of this theory are
precisely those given by the independent parameters of a WZW model with target
space the subgroup $P$ of $G$ in ref. [\us].  Finally, it is straightforward to
construct a map between the space of parameters of the solutions  and the space
of initial data of the non-abelian Toda theories, and prove that it is an
isomorphism of symplectic spaces.

%%%%%%%%%%%%%%%%%%%%%%%%%%%%%%%%%%%%%%%%%%%%%%%%%%%%%%%%%%%%%%%%%%%%%%%%%%%

\chapter {Concluding Remarks}

The parameterisation described in section three can also be used to study
Toda theory on a two-dimensional Minkowski space.  Indeed,  a parameterisation
of the solutions of the Toda theory is given by eqn. \tone\ but in this case
the parameters $a$ and $\phi_L$ are not necessarily periodic functions on the
real line.  A similar conclusion can be drawn for the non-Abelian Toda
theories.  The calculation of the symplectic form and the Poisson brackets of
the theory can be formally described as in section four.

To summarise, we have presented a new parameterisation of the space of
solutions of Toda theory on a cylinder.  In this parameterisation, the
solutions of the theory are parameterised by two sets of parameters, one of
which is a set of components of a connection.  This connection appeared in the
expression for the solutions via its holonomy. It is worth pointing out that
the holonomy of a connection entered in the parameterisation of the solutions
of Liouville and WZW models [\usthree, \us] as well.  Thus we have found a
unified and geometric understanding of the parameterisations of the solutions
of the WZW, Liouville and Toda field theories.    We  also showed that the
covariant phase space of Toda theory is isomorphic to the co-tangent bundle of
the loop space of $\RN^l$, with the associated Poisson brackets.  In addition
we constructed a diffeomorphism from the space of parameters of the solutions
of Toda theory to the space of initial data of the theory, showed that the
solutions are well-defined and real functions over all of space-time, and
proved that the Hamiltonian and covariant phase spaces of Toda theory are
diffeomorphic as symplectic manifolds.
Finally, we discussed how these results can be extended to include the models
that are obtained from the WZW model via non-standard Hamiltonian reduction
methods (the non-Abelian Toda theories).

\noindent{\bf Acknowledgements:} We would like to thank L. Feh\'er for
discussions.  G.P. was supported by the Commission of
European Communities, and B.S. by a QEII Fellowship from the Australian
Government.

\refout
\end